# Deployable Nanoelectromechanical Bound States in the Continuum Enabled by GHz Lamb Wave Phononic Crystals on LiNbO₃ Thin Films

Sheng-Nan Liang[1†], Zhen-Hui Qin[1†], Shu-Mao Wu[1], Hua-Yang Chen[1],
Si-Yuan Yu[1,2,3]* and Yan-Feng Chen[1,2,3]*

[1]National Laboratory of Solid-State Microstructures & Department of Materials Science and Engineering, Nanjing University, Nanjing 210093, China.
[2]Collaborative Innovation Center of Advanced Microstructures, Nanjing University, Nanjing 210093, China
[3]Jiangsu Key Laboratory of Artificial Functional Materials, Nanjing University, Nanjing, 210093, China
† These authors contributed equally: Sheng-Nan Liang, Zhen-Hui Qin
*corresponding author. e-mail: yusiyuan@nju.edu.cn; yfchen@nju.edu.cn

**Bound states in the continuum (BICs) are a fascinating class of eigenstates that trap energy within the continuum, enabling breakthroughs in ultra-low-threshold lasing, high-Q sensing, and advanced wave-matter interactions. However, their stringent symmetry requirements hinder practical integration, especially in acoustic and electromechanical systems where efficient mode excitation is challenging. Here, we demonstrate deployable nanoelectromechanical quasi-BICs on suspended lithium niobate (LiNbO₃) thin films, enabled by nanoscale Lamb wave phononic crystals (PnCs) operating at gigahertz frequencies. By exploiting the decoupling of symmetric ($S$) and antisymmetric ($A$) Lamb wave modes, we create a robust framework for BICs. Controlled mirror symmetry breaking induces targeted coupling between the $S$ and $A$ modes, resulting in quasi-BICs that preserve high-$Q$ characteristics and can be excited by traveling waves, eliminating the need for specialized excitation schemes. Our approach enables the multiplexing of quasi-BIC resonators along a single transmission line, each corresponding to a unique frequency and spatial position. This work presents a scalable route for the on-chip integration of BICs, bridging the gap between theoretical concepts and practical nanoelectromechanical devices, and opening new avenues in advanced signal processing, high-precision sensing, and quantum acoustics.**

Bound states in the continuum (BICs) are a class of remarkable eigenstates that confine energy within the radiative continuum, enabling advances in applications such as ultra-low-threshold lasing, high-$Q$ sensing, and novel wave-matter interactions. Originally proposed in quantum mechanics [1,2], BICs have since been explored extensively across various fields, including photonics [3-14] and acoustic systems [15-25]. In the context of acoustics, BICs were first demonstrated in airborne sound, leveraging their clean longitudinal wave modes and efficient waveguiding and localization. More recently, efforts have focused on extending BICs to solid-state acoustic systems, where shear wave interactions and mode hybridization introduce significant challenges [26-31]. Despite these challenges, BICs in solid-state systems, particularly in the realm of surface acoustic waves, have shown great promise for applications in high-precision sensing and signal processing [27,31]. However, achieving practical device integration of BICs in acoustic and electromechanical systems remains a major hurdle, largely due to the stringent symmetry requirements and difficulties in efficiently exciting the modes.

In this Letter, we present the first experimental realization of deployable nanoelectromechanical quasi-BICs on suspended lithium niobate (LiNbO₃) thin films, enabled by nanoscale Lamb wave phononic crystals (PnCs) operating at gigahertz frequencies. By leveraging the decoupling between symmetric ($S$) and antisymmetric ($A$) Lamb wave modes, we develop a robust framework for constructing mechanical BICs. Controlled mirror symmetry breaking induces targeted coupling between the $S$ and $A$ modes, resulting in quasi-BICs that preserve high-$Q$ characteristics while allowing direct traveling-wave excitation, eliminating the need for specialized in-situ excitation schemes. Our findings introduce a scalable approach for integrating quasi-BIC resonators into a single transmission line, unveiling new opportunities for multi-band filtering, multi-mode sensing, and integrated wave-matter interaction, while bridging the divide between theoretical models and practical, chip-scale acoustic technologies.

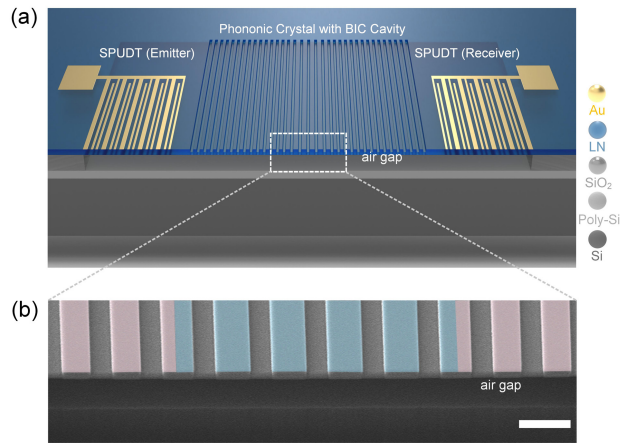

FIG. 1. (a) Schematic of the deployable BIC resonator on a 30° X-cut LiNbO₃ thin film. SPUDTs flanking the nanoscale PnC generate and detect the BIC resonator at the center. (b) SEM image of the fabricated LiNbO₃ thin film with periodic groove etchings. The blue region has a larger period/width than the pink region, forming the BIC. Scale bar: 2 μm.

**Realization of Deployable BIC on Suspended LiNbO₃ Thin Films**

We successfully realize deployable BIC resonators on a suspended LiNbO₃ transmission line (710 nm thick, X-cut) oriented along the 30° Y-crystal direction (Fig. 1(a)). One-dimensional (1D) Lamb wave PnCs are fabricated by etching narrow grooves on the surface of the LiNbO₃ film, with periodic defects introduced at the center to form the BIC. These grooves are designed to facilitate the excitation of the symmetric ($S$) mode of the Lamb wave, which exhibits the highest electromechanical coupling coefficient, ensuring efficient excitation via piezoelectric transducers at both ends of the transmission line.

In our experiments, two single-phase uni-directional transducers (SPUDTs) were used: one as the emitter and the other as the receiver. This setup allowed us to confirm the presence of the quasi-BIC through transmission-line measurements. Fig. 1(b) presents an SEM image of the fabricated LiNbO₃ thin film with periodic groove etchings, where the blue region has a larger period and width compared to the pink region, forming the BIC. The fabrication process involves (i) dry-etching the LiNbO₃ layer using patterned silicon and chromium masks, (ii) selectively etching and removing the silicon dioxide (SiO₂) buffer layer using BOE solution, and (iii) achieving a fully suspended LiNbO₃ thin film through critical-point drying. Further details of the fabrication process are provided in the Methods section and Supplementary Information (SI, Section I).

**Enabling Mechanical BICs and Quasi-BICs by Lamb Wave PnCs**

Fig. 2 outlines the principle and methodology for constructing BICs and quasi-BICs in plate-wave PnCs, starting with the simplest case of a free plate. The Rayleigh–Lamb equation governs the dispersion relation for acoustic waves propagating through such plates [32]:

$$\frac{\tan(K_T h)}{\tan(K_L h)} - \left[\frac{4k^2 K_T K_L}{(K_T^2 - K_L^2)^2}\right]^{\pm 1} = 0$$

Where $K_L^2 = k_L^2 - k^2$, $K_T^2 = k_T^2 - k^2$, with $K_L$ and $K_T$ representing the longitudinal and transverse wave components along the z-axis. The wave numbers $k_L = \omega/c_L$ and $k_T = \omega/c_T$ are defined in terms of the longitudinal and transverse wave speeds.

In a free plate, the acoustic waves are decoupled into two distinct modes — symmetric ($S$) and antisymmetric ($A$) — based on their displacement distributions relative to the mid-plane (z = 0). Fig. 2(a) shows the dispersion spectra of the $S$ and $A$ modes for a free plate. These modes are independent, and their dispersion curves overlap without mutual interference.

To achieve mode separation, we introduce PnCs by etching periodic grooves that maintain the mid-plane symmetry ($M_z$) of the plate. Fig. 2(b) shows the resulting band gaps for both the $S$ and $A$ modes, successfully separating them in frequency. The grooves' positions, width, and depth determine the exact location of the band gaps. Importantly, in this $M_z$-symmetric PnC, the $S$ and $A$ modes remain decoupled, preserving their character as in the free plate. The insets further confirm that the displacement fields of the modes retain the mid-plane symmetry.

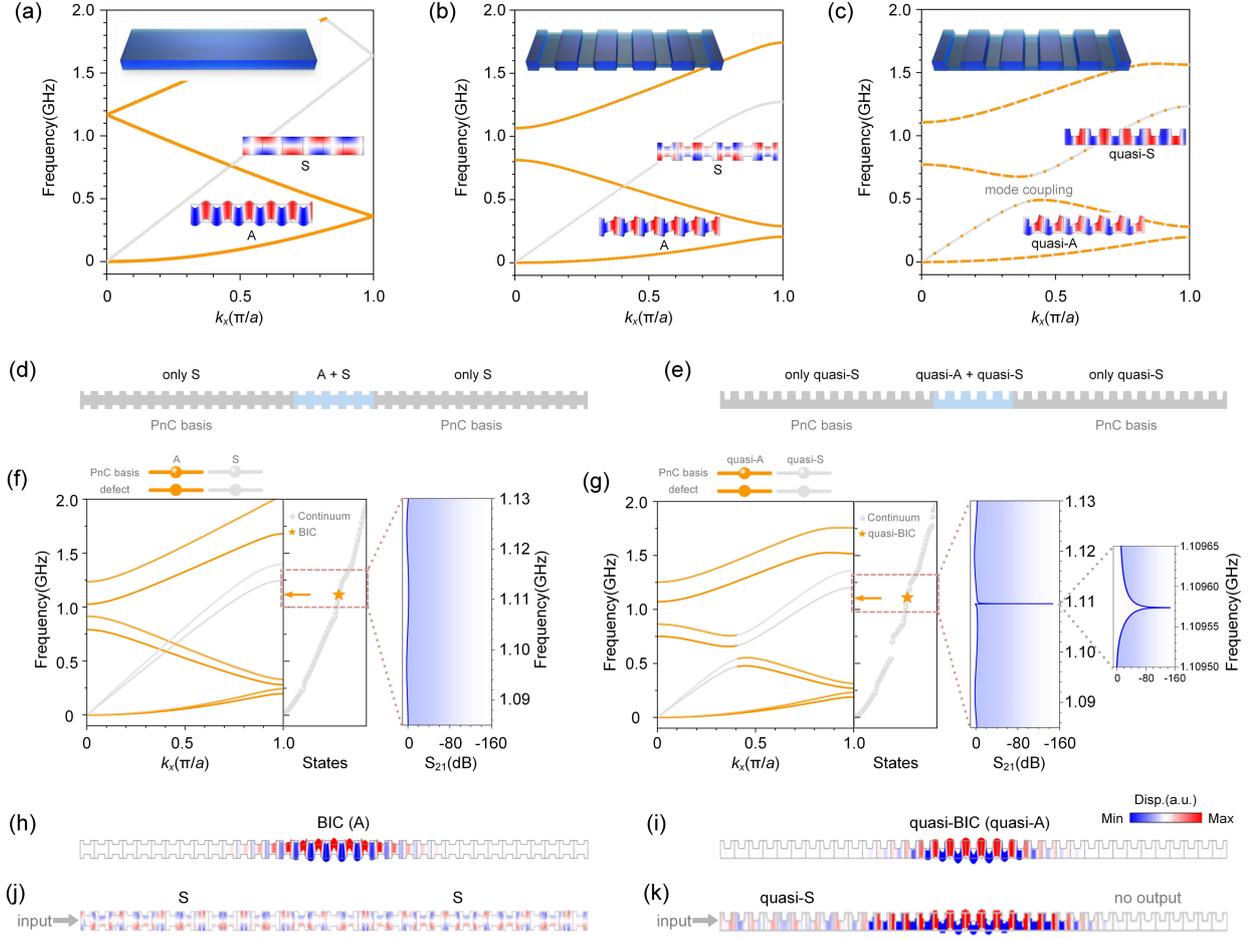

FIG. 2. (a) Folded dispersion diagram and eigenmodes (*S* and *A*) of a free plate. (b) Dispersion diagram and eigenmodes of a $M_z$-symmetric plate PnC, with grooves etched on both the upper and lower surfaces. (c) Dispersion diagram and eigenmodes of a $M_z$-asymmetric plate PnC, where grooves are etched on only one surface, causing coupling between the *S* and *A* modes (quasi-*S* and quasi-*A*). (d),(e) BIC resontaors constructed by the $M_z$-symmetric (d) and $M_z$-asymmetric (e) plate PnCs, respectively. (f), (g) (Left) Dispersion diagrams for the PnC basis and defect regions; (Middle) mode solution spectra illustrating the emergence of BICs and quasi-BICs; Right: calculated two-port transmission ($S_{21}$) for the corresponding resontaors. (h),(i) Eigenmode-calculated in-plane displacement fields for the BIC (h) and quasi-BIC (i). (j),(k) In-plane displacement fields under traveling-wave input for the $M_z$-symmetric (j) and $M_z$-asymmetric (k) PnCs, showing no excitation output for the BIC versus strong coupling for the quasi-BIC.

Fig. 2(d) illustrates how a 1D PnC can be used to form a strict BIC by combining a periodic PnC "basis" with a defect region at the center. The lattice constant in the defect is slightly larger than in the PnC basis, and within the targeted frequency range, the PnC basis supports only the continuous *S* mode, while the *A* mode is confined to the defect region. This results in the formation of a strict BIC, as shown by the eigenmode solution in Fig. 2(f), with the field distribution shown in Fig. 2(h). Because the PnC is $M_z$-symmetric, the *A*-mode BIC cannot be excited by traveling *S* modes in the surrounding

PnC, as confirmed by the two-port transmission ($S_{21}$) remaining near unity for the entire *S*-mode band (Fig. 2(j)).

To enable coupling between the *S* and *A* modes, we break the $M_z$ symmetry by etching grooves only on one surface of the plate (Fig. 2(c)). This results in the splitting of the dispersion curves for the *S* and *A* modes, creating a frequency gap where both modes coexist, now referred to as quasi-*S* and quasi-*A* modes. Fig. 2(e) shows that in this asymmetric PnC design, a quasi-BIC can be formed. Here, only the top side of the plate is etched, and within the continuous passband (carried by the quasi-*S* mode), a confined quasi-*A* mode BIC emerges (Fig. 2(g)). Because the *S* and *A* modes are no longer strictly independent, this BIC is a quasi-BIC with a finite *Q* factor, as confirmed by the $S_{21}$ transmission dip at the quasi-BIC frequency (Fig. 2(k)).

**Experimental Verification of Quasi-BICs in Integrated GHz LiNbO₃ Transmission Lines**

To experimentally verify the quasi-BIC, we fabricated 1D PnCs with embedded BIC resonators on a 710 nm-thick X-cut LiNbO₃ thin film. These devices were integrated into suspended transmission lines with SPUDTs for excitation and reception. Fig. 3(a) and 3(b) show optical and SEM images of the fabricated device, where the core functional region is fully suspended. The PnC consists of 35 periodic grooves, and the defect region includes five substitutional defects.

Figs. 3(c) and 3(d) illustrate the specific geometries of the substitutional defects in Sample #1 and Sample #2, respectively. These defects are crucial for forming the quasi-BIC resonators at the center of the PnC. Transmission spectra measured for these samples (Figs. 3(e) and 3(f)) show prominent dips around 1.1 GHz, confirming the presence of quasi-BICs. These dips are consistent with theoretical predictions and indicate that traveling quasi-S waves are converted into quasi-*A* (quasi-BIC) modes within the defect regions.

Additionally, we performed low-temperature measurements to investigate the temperature dependence of the quasi-BIC. Fig. 3(g) shows the $S_{21}$ spectra for one of the samples at temperatures ranging from 300 K to 80 K. As the temperature decreases, the quasi-BIC frequency shifts slightly from 1.1059 GHz at 300 K to 1.1096 GHz at 80 K, due to thermal expansion and variations in the elastic properties of LiNbO₃. Furthermore, the *Q* factor increases from ~4700 at 300 K to ~7800 at 80 K, as shown in Fig. 3(h), indicating reduced thermoelastic losses at lower temperatures. The inset in Fig. 3(h) shows the ring-down measurement at room temperature, revealing a lifetime of $\tau \approx 480$ ns, corresponding to a *Q* factor of approximately 3500.

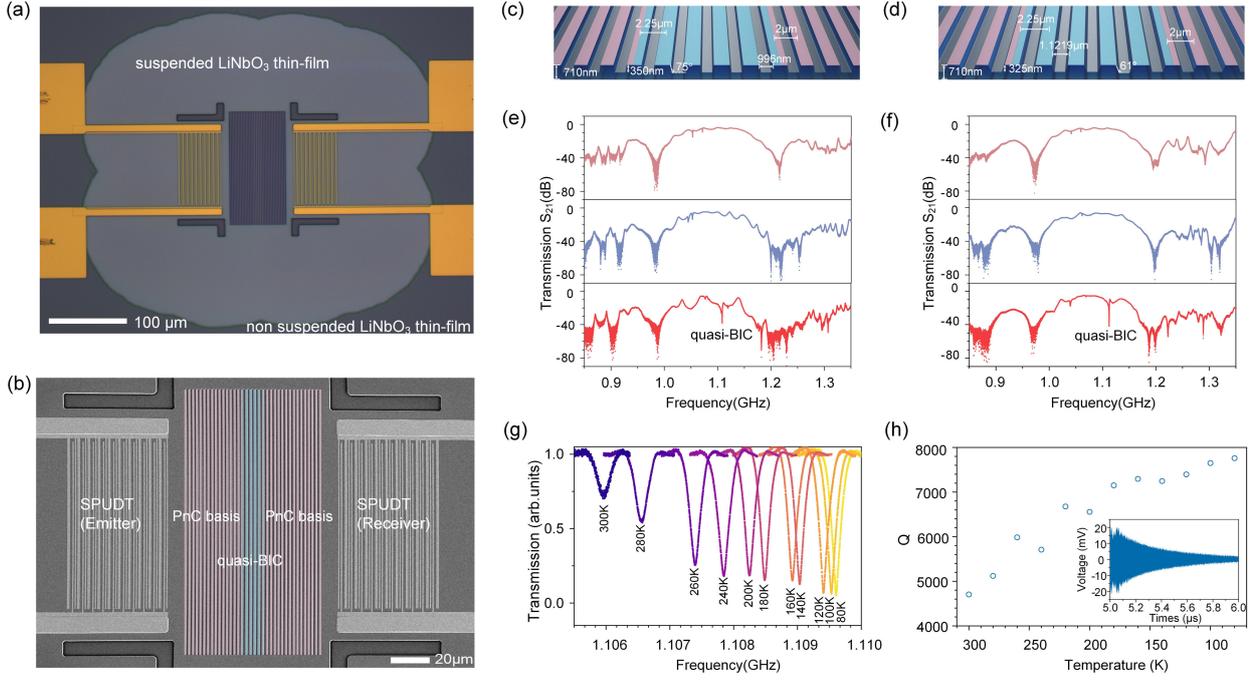

FIG. 3. (a),(b) Optical microscope and SEM images of a fabricated device, which features SPUDTs on both sides of a 35-period PnC basis containing an embedded cavity. The core structure is formed on a fully suspended, 710 nm-thick LiNbO$_3$ thin film. (c),(d) Samples #1 and #2, illustrating the specific geometries of the substitutional defects in the PnC center. (e),(f) Measured two-port transmission ($S_{21}$) spectra for Samples #1 and #2, respectively. Top: transmission lines with only SPUDTs; middle: lines with the PnC basis (no defects); bottom: line with both the PnC basis and defects. Prominent dips around 1.1 GHz confirm the quasi-BICs. (g) Low-temperature $S_{21}$ spectra of another sample's quasi-BIC, measured from 300 K down to 80 K. (h) Temperature-dependent $Q$ factor of the quasi-BIC. Inset: Ring-down measurements at room temperature, illustrating the device's high-$Q$ resonance.

**Deploying Multiple Quasi-BICs in a Single Transmission Line**

In contrast to traditional bandgap-confined states, quasi-BICs can be excited by passband modes, allowing multiple quasi-BICs to coexist along the same transmission line without degrading each other's performance. This design flexibility enables the positioning of quasi-BICs at different spatial locations along the transmission line, each corresponding to a distinct resonant frequency.

Fig. 4(a) shows an SEM image of a fabricated device containing three quasi-BICs, each with a different geometry. The three quasi-BICs are excited at frequencies of 1.145 GHz, 1.150 GHz, and 1.155 GHz, respectively. Figs. 4(b)-4(d) show simulated energy field distributions and out-of-plane displacement profiles for the three quasi-BICs under the same input power. Despite their different spatial locations, each quasi-BIC receives comparable excitation intensity.

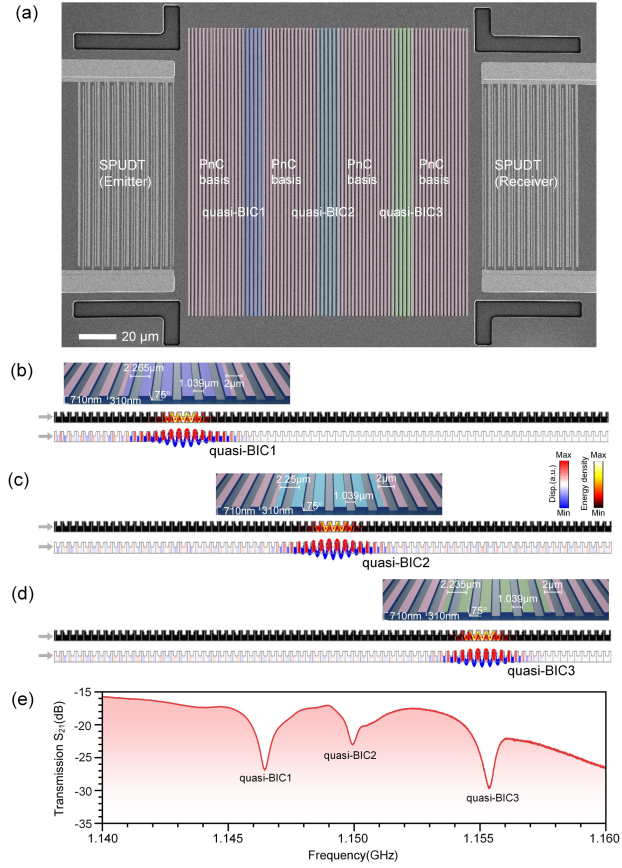

FIG.4 (a). SEM images of the fabricated devices, showing three quasi-BICs, each with distinct geometries. (b), (c), (d) Simulated energy field distributions and out-of-plane displacement profiles for the three quasi-BICs excited at 1.145 GHz, 1.150 GHz, and 1.155 GHz, respectively, under the same input power. Despite their different positions, each quasi-BIC experiences comparable excitation intensity. (e) Experimentally measured transmission spectra, revealing three discrete dips corresponding to the quasi-BICs.

Experimental measurements (Fig. 4(e)) reveal three distinct dips in the $S_{21}$ transmission spectra, corresponding to the three quasi-BICs. These results demonstrate the feasibility of multiplexing multiple quasi-BICs in a single transmission line, providing new opportunities for multi-band filtering, multi-mode sensing, and enhanced wave-matter interactions in a compact, integrated platform.

**Conclusion**

This work presents the first experimental demonstration of deployable nanoelectromechanical quasi-BICs on suspended $LiNbO_3$ thin films, enabled by nanoscale PnCs operating at gigahertz frequencies. Through controlled mirror symmetry breaking, we achieve direct coupling between symmetric and antisymmetric Lamb wave modes, resulting in quasi-BICs with high-$Q$ characteristics that can be efficiently excited by traveling waves. This innovation eliminates the need for specialized in-situ excitation techniques and enables the deployment of BICs without positional constraints within the PnC. We demonstrate that multiple quasi-BICs with distinct frequencies can coexist along a single transmission line with comparable excitation efficiency, offering significant advantages over conventional bandgap-confined states. This work opens new avenues for multi-band filtering, multi-mode sensing, and integrated wave-matter interactions, all within a compact, on-chip platform. The combination of LiNbO$_3$'s outstanding material properties with our PnC design ensures a wide operational range, from room temperature to cryogenic conditions. Our findings lay the groundwork for scalable, high-performance nanoelectromechanical devices, advancing the practical implementation of BICs in real-world acoustic technologies.